\documentclass[aps,prl,twocolumn,superscriptaddress,floatfix]{revtex4-2}

\usepackage{amsmath}
\usepackage{graphicx}
\usepackage{color}

\begin{document}

\title{Direct view of exciton and carrier dynamics in twisted bilayer graphene}
\author{Gianmarco Gatti}
\thanks{These authors contributed equally to this work.}
\affiliation{Department of Physics and Astronomy, Aarhus University, 8000 Aarhus C, Denmark}
\author{Anders S. Mortensen}
\thanks{These authors contributed equally to this work.}
\affiliation{Department of Physics and Astronomy, Aarhus University, 8000 Aarhus C, Denmark}
\author{Zhihao Jiang}
\affiliation{Department of Physics and Astronomy, Aarhus University, 8000 Aarhus C, Denmark}
\author{Alfred J. H. Jones}
\affiliation{Department of Physics and Astronomy, Aarhus University, 8000 Aarhus C, Denmark}
\author{Gesa-R. Siemann}
\affiliation{Department of Physics and Astronomy, Aarhus University, 8000 Aarhus C, Denmark}
\author{Thomas S. Nielsen}
\affiliation{Department of Physics and Astronomy, Aarhus University, 8000 Aarhus C, Denmark}
\author{S\o ren Ulstrup}
\affiliation{Department of Physics and Astronomy, Aarhus University, 8000 Aarhus C, Denmark}
\author{Naina Kushwaha}
\affiliation {Central Laser Facility, STFC Rutherford Appleton Laboratory, OX11 0QX, Harwell, UK}
\affiliation{SUPA, School of Physics and Astronomy, University of St Andrews, North Haugh, KY16 9SS, St Andrews, UK}
\author{Bruce Weaver}
\affiliation {Central Laser Facility, STFC Rutherford Appleton Laboratory, OX11 0QX, Harwell, UK}
\author{Yu Zhang}
\affiliation {Central Laser Facility, STFC Rutherford Appleton Laboratory, OX11 0QX, Harwell, UK}
\author{Emma Springate}
\affiliation {Central Laser Facility, STFC Rutherford Appleton Laboratory, OX11 0QX, Harwell, UK}
\author{Charlotte E. Sanders}
\affiliation {Central Laser Facility, STFC Rutherford Appleton Laboratory, OX11 0QX, Harwell, UK}
\affiliation{SUPA, School of Physics and Astronomy, University of St Andrews, North Haugh, KY16 9SS, St Andrews, UK}
\author{Vibha Reddy}
\affiliation{Max Planck Institute for Solid State Research, 70569 Stuttgart, Germany}
\author{Ulrich Starke}
\affiliation{Max Planck Institute for Solid State Research, 70569 Stuttgart, Germany}
\author{Philip Hofmann}
\email{philip@phys.au.dk}
\affiliation{Department of Physics and Astronomy, Aarhus University, 8000 Aarhus C, Denmark}
\date{\today}

\begin{abstract}

Twisted bilayer graphene (tBLG) on H-intercalated SiC (twist angle of $9 \pm 1^{\circ}$) is studied by position\mbox{-}, time\mbox{-}, and angle-resolved photoemission spectroscopy (ARPES). Both graphene layers are hole doped, with a Dirac-point offset of 30~meV. Time-resolved ARPES near the van Hove singularity reveals a spectral feature within the hybridization-induced band gap, together with excited electrons in the conduction band. The in-gap feature exhibits distinct dynamics, and its binding energy and $k$-space location are consistent with the proposed non-hydrogenic ``ghost-Fano'' exciton arising from destructive interference between two nearly degenerate excitonic channels. Near the Dirac points, the twist separates the largely layer-polarized Dirac cones in momentum space, providing direct layer selectivity. We find markedly different peak electronic temperatures in the two graphene layers, even though their subsequent cooling dynamics are very similar. We attribute this to an ultrafast interlayer redistribution of electronic energy, driven by an interlayer electric field that arises from the different doping in the layers.
\end{abstract}

\maketitle

Twisted bilayer graphene (tBLG) exhibits a variety of physical phenomena near the ``magic'' twist angle of $\approx$~1.1$^{\circ}$, where the band structure in the mini Brillouin zone (BZ) created by the moir\'e pattern consists of flat bands with electrically controllable filling \cite{Bistritzer:2011aa,Cao:2018aa,Cao:2018ad}. At larger twist angles, the hybridization of the two Dirac cones occurs away from the Fermi energy and produces van Hove singularities (vHs) in the electronic and joint density of states \cite{Li:2009ab,Brihuega:2012aa,Ohta:2012aa,Havener:2012aa,Moon:2013aa,Tabert:2013aa,Havener:2014aa}. The energy of these vHs is tunable by twist angle and by an external electric field \cite{Jones:2020aa}. 

The twist-angle-controllable interband transitions near the vHs have been predicted to host unusual excitonic states \cite{Liang:2014aa}. While excitonic effects are known to be important in graphene \cite{Yang:2009aa,Mak:2011aa}, in tBLG the band splitting near the vHs gives rise to two nearly degenerate excitonic channels, each formed between two of the four bands making up the upper and lower vHs. These excitons can interact both with the continuum of electron-hole excitations and with each other. This has been predicted to produce a ``ghost-Fano'' resonance \cite{Guevara:2003aa,Liang:2014aa}, comprising a bright resonant exciton and a stable, strongly bound dark exciton. Signatures consistent with the dark exciton have been reported in optical experiments \cite{Patel:2015aa,Patel:2019aa}, but its microscopic character has not been directly established.  Unlike a hydrogenic exciton, the ghost-Fano exciton can remain strongly bound with a nm-scale spatial extent even in a metallic background \cite{Liang:2014aa}. With its momentum resolution, time- and angle-resolved photoemission spectroscopy (TR-ARPES) is well suited for determining the $k$-space localization and spatial extent of such excitons \cite{Weinelt:2004aa,Madeo:2020aa,Dong:2021um}.

In this Letter, we use spatially resolved TR-ARPES to study the non-equilibrium electronic structure of large-angle tBLG. Near the hybridization-induced vHs, we probe the predicted excitonic effects. The large momentum separation of the layer-polarized states near the Dirac points also allows us to follow the redistribution and relaxation of electronic energy between the individual layers of a twisted graphene homo-bilayer.

tBLG samples were fabricated by polymer-assisted transfer of single-layer graphene flakes onto H-intercalated quasi-free-standing monolayer graphene (QFMLG) on SiC \cite{Riedl:2009aa}. In this process, the twist angle can be chosen arbitrarily, and the typical size of the resulting tBLG islands is tens of $\mu$m. The equilibrium electronic structure of the flakes was characterized using the SGM-4 microARPES beamline at ASTRID2 \cite{Jones:2025aa}. The sample temperature was 300~K, the spatial resolution was $\approx$~3~$\mu$m, the photon energy was 70~eV, and the energy and $k$ resolutions were 50~meV and 0.02~\AA$^{-1}$, respectively. Time-resolved photoemission experiments were carried out at the Artemis laboratory of the Central Laser Facility. Pump and probe beams were generated from the same 100~kHz laser system~\cite{Thire:2023aa}, with their respective linear polarizations perpendicular and parallel to the plane of incidence. The pump and probe photon energies were 1.6 and 37~eV, respectively; the time resolution was $\approx$~80~fs, the energy resolution was better than 150~meV, and the $k$ resolution along the analyzer slit was better than 0.015~\AA$^{-1}$. Temporal overlap between pump and probe beam ($\Delta t = 0$) was determined from the fit shown in Fig.~\ref{fig:2}(d). The pump fluence was set to 0.75~mJ/cm$^2$. Tight focusing of the probe beam and spatial scanning across the sample enabled measurements with simultaneous time and spatial resolution. The spatial resolution was 10~$\mu$m, and the sample temperature was 300~K. The band structure of the tBLG was calculated using a simple tight-binding model, with parameters chosen to reproduce the experimental dispersion. Further details on the experiments and calculations are given in the Appendix.

\begin{figure}[t]
\includegraphics[width=0.5\textwidth]{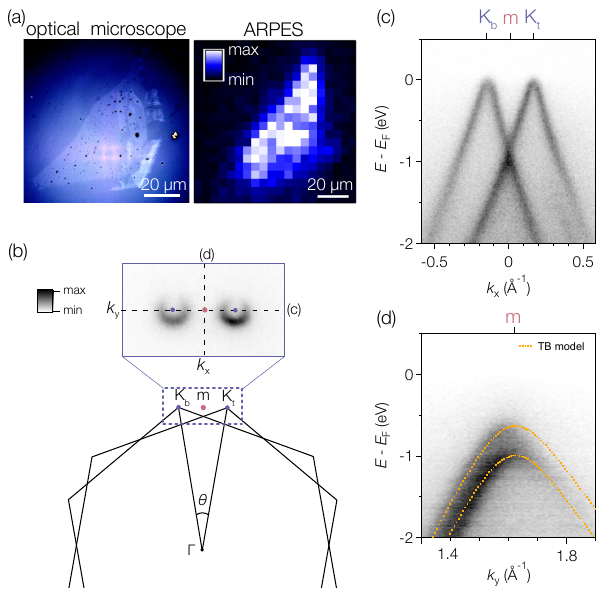}
\caption{(Color online) Characterization of the tBLG sample. 
(a) Optical image and photoemission intensity in an $(E,k)$ region of interest around the top-layer Dirac cone, measured at the Artemis TR-ARPES facility. 
(b) Photoemission intensity near the Brillouin-zone corner 300~meV below the Fermi energy, together with a sketch of the two Brillouin zones. 
(c), (d) Photoemission intensity along the two dashed lines in panel (b), showing the electronic dispersion near the van Hove singularity. The dotted orange lines in panel (d) show the band structure calculated using a tight-binding model. The data in panels (b)--(d) were obtained at the SGM-4 microARPES setup. 
}
\label{fig:1}
\end{figure}

The tBLG island is characterized in Fig.~\ref{fig:1}. Panel (a) shows an optical image of the transferred flake together with a map of the photoemission intensity in a region of interest around the top-layer Dirac cone. The map was collected at Artemis, illustrating the spatial resolution of the TR-ARPES setup. The rest of the figure shows data collected at SGM-4. Fig.~\ref{fig:1}(b) shows the photoemission intensity 300~meV below the Fermi energy near the K points of the two graphene layers, together with the two twisted BZs. The circular contours, arising from the $\pi$ bands of the two layers, define the K points of the bottom and top layers, denoted K$_\mathrm{b}$ and K$_\mathrm{t}$, respectively. From wider constant-energy maps, we determine a twist angle of $\theta = 9\pm 1^{\circ}$ , see Appendix.  The top-layer contour is more intense and slightly smaller, indicating lower hole doping. The difference in doping between the two layers can be estimated from the energy positions of the Dirac points, obtained by fitting the linearly dispersing bands. They are located 63$\pm 9$ and 93$\pm 7$~meV above $E_{\mathrm{F}}$ for the top and bottom layers, respectively. 
This is consistent with previous results \cite{Ohta:2012aa} and indicates a carrier density difference of $\approx (3\pm 1)\times10^{11}$~cm$^{-2}$. This difference is attributed to the bottom-layer doping being fixed by the substrate, while the top layer is partly screened from the substrate by the bottom layer and is therefore closer to charge neutrality. 

Figures~\ref{fig:1}(c) and \ref{fig:1}(d) show the low-energy dispersion along the dashed lines in Fig.~\ref{fig:1}(b), corresponding to the line connecting K$_\mathrm{b}$ and K$_\mathrm{t}$ and to a radial cut from $\Gamma$ through the midpoint between K$_\mathrm{b}$ and K$_\mathrm{t}$, respectively. The avoided crossing of the two Dirac cones produces a vHs at the m point of the mini BZ, marked in Fig.~\ref{fig:1}(b). This feature is clearly visible in Fig.~\ref{fig:1}(d), while it is less pronounced in Fig.~\ref{fig:1}(c) \cite{Ohta:2012aa}. The combination with a corresponding vHs in the unoccupied states gives rise to a critical point in the joint density of states \cite{Li:2009ab,Brihuega:2012aa,Ohta:2012aa,Havener:2012aa,Moon:2013aa,Tabert:2013aa,Havener:2014aa}. We fit a tight-binding model to the observed band structure and overlay the result as dotted lines in Fig.~\ref{fig:1}(d) (see Appendix).

\begin{figure}[t]
\includegraphics[width=0.5\textwidth]{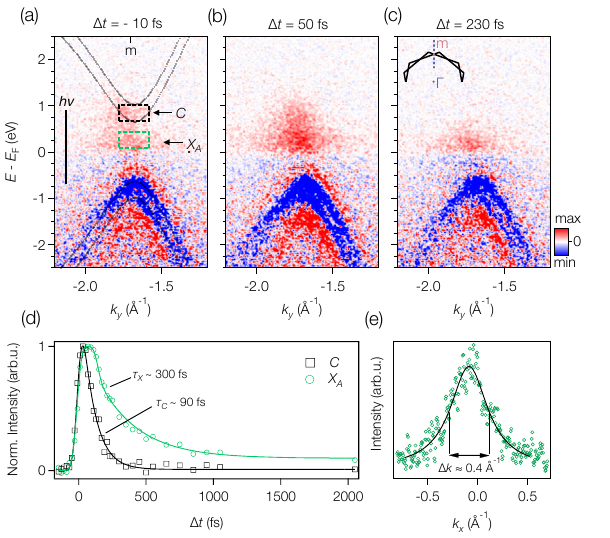}
\caption{(Color online) Excited electron and exciton dynamics at the m point of the mini Brillouin zone, where the vHs is located, in 9$^{\circ}$ tBLG.
(a)--(c) Photoemission intensity difference plots obtained by subtracting an ARPES spectrum measured 250~fs before the temporal overlap $t_0$ of the pump and probe pulses from spectra measured at selected time delays $\Delta t = t - t_0$. The data were measured along the same momentum path as in Fig.~\ref{fig:1}(d). In panel (a), the tight-binding band structure is overlaid as a black dotted line. A vertical bar indicates the pump photon energy.
(d) Normalized photoemission intensity integrated over the dashed boxes of the same color in panel (a). Solid lines are fits with a function including pulsed excitation and a single-exponential decay.  
(e) Momentum distribution curve at the center energy of the $X_A$ feature (0.25~eV) and $\Delta t =-10$~fs, together with a Gaussian fit.
}
\label{fig:2}
\end{figure}

We next investigate the electron dynamics at the m point, where the vHs is located. Figs.~\ref{fig:2}(a)--\ref{fig:2}(c) show snapshots of the photoemission intensity difference, obtained by subtracting the equilibrium spectrum from spectra measured at different pump-probe delays $\Delta t$, for the same cut as in Fig.~\ref{fig:1}(d). Two distinct features are visible, most clearly at early delays around $\Delta t = -10$~fs. These are labeled $C$ and $X_A$. Fig.~\ref{fig:2}(d) shows the normalized photoemission intensity integrated over energy-momentum regions enclosing these features. The two signals exhibit markedly different dynamics: $X_A$ shows a broad maximum, decays more slowly, and remains visible at longer delays. Comparison with the tight-binding band structure overlaid in Fig.~\ref{fig:2}(a) assigns the $C$ feature to the conduction band minimum, although the conduction band splitting is not resolved. Particle--hole asymmetry in this energy range is sufficiently small that it does not affect this assignment~\cite{Brihuega:2012aa}.

\begin{figure}[t]
\includegraphics[width=0.5\textwidth]{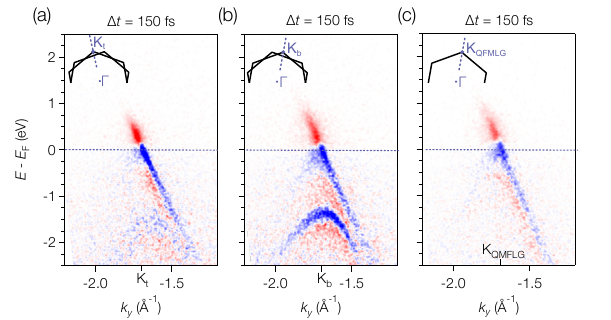}
\caption{(Color online) Excited carrier dynamics at the Dirac cones of the top layer, bottom layer, and bare QFMLG.  
Difference plots at $\Delta t=150$~fs near (a) K$_\mathrm{t}$, (b) K$_\mathrm{b}$, and (c) K$_\mathrm{QFMLG}$, where K$_\mathrm{QFMLG}$ denotes the K point of the uncovered quasi-free-standing monolayer graphene on SiC.
}
\label{fig:3}
\end{figure}

With this assignment, the $X_A$ structure lies inside the hybridization-induced gap, approximately $0.4\pm 0.1$~eV below the tight-binding conduction band minimum. We assign this feature to the strongly bound ghost-Fano exciton arising from the antisymmetric coherent superposition of the multiband excitons predicted in Ref.~\cite{Liang:2014aa}. Several experimental observations support this assignment. First, the energy separation of $\approx 0.4$~eV between the $X_A$ feature and the  conduction band feature is of the same order of magnitude as the predicted binding energy. This separation provides only an approximate estimate of the binding energy, since the coherent superposition of two interband excitations may in principle produce two residual-hole contributions to the $X_A$ photoemission signal~\cite{Christiansen:2019aa,Bange:2024aa,Bennecke:2024aa}. Such a splitting is not resolved here, consistent with the unresolved conduction band splitting and the experimental energy resolution. Second, the $X_A$ feature is confined to the vicinity of the m point. It dominates the difference intensity in this region but is not observed in the corresponding TR-ARPES data near K$_\mathrm{b}$ and K$_\mathrm{t}$, as shown in Fig.~\ref{fig:3}. Its width in momentum space can be used to estimate the exciton size in real space~\cite{Rustagi:2018aa,Dong:2021um,Man:2021aa}. Fig.~\ref{fig:2}(e) shows a Gaussian fit to the momentum distribution of the $X_A$ feature at $\Delta t=-10$~fs and $E-E_\mathrm{F}=0.25$~eV, yielding a full width at half maximum (FWHM) of $\approx 0.4$~\AA$^{-1}$.
Assuming a Gaussian envelope, this corresponds to a real-space exciton extent (FWHM) of $\approx 4\ln(2)/(0.4~\text{\AA}^{-1})=0.7$~nm, within the expected order of magnitude for the strongly bound exciton~\cite{Liang:2014aa,Patel:2015aa,Patel:2019aa}.

Third, the $X_A$ signal shows a relatively slow decay, with a time constant of $\tau_X=300$~fs obtained from the fit in Fig.~\ref{fig:2}(d). While this decay time is shorter than the $\sim 1$~ps excitonic relaxation time inferred from transient-absorption measurements in Ref.~\cite{Patel:2019aa}, this does not necessarily imply a contradiction, since the latter is an indirect optical measurement, and the reported time includes population transfer into $X_A$ and the subsequent decay. In contrast, we follow the photoemission intensity assigned to $X_A$ and thus obtain a direct estimate of the photoemission-visible exciton population and its lifetime. A lifetime of several hundred fs is remarkable for a gapless semimetal. In conventional metals, excitonic correlations usually survive only during the few-femtosecond buildup of screening~\cite{Cui:2014aa}. Although this dephasing time is not directly equivalent to our population decay, the large difference in timescale is consistent with the ghost-Fano mechanism suppressing coupling to the electron--hole continuum. Finally, the $X_A$ signal exhibits a plateau around its maximum, consistent with continued population from higher-lying states after the initial excitation. Such an indirect population pathway is expected because direct optical excitation of $X_A$ is dipole-forbidden and we use and above-gap pump energy.

We now turn to the excited-carrier dynamics near the Dirac points. For the large twist angle studied here, these states remain largely layer-polarized, allowing us to determine the electronic temperature $T_e(t)$ of the top and bottom graphene layers separately. Fig.~\ref{fig:3} shows the hot-carrier distributions in the top layer, bottom layer, and, as a reference, bare QFMLG shortly after peak excitation. In panels (a) and (b), excitation from the lower-lying band near the vHs is also visible, but this region of interlayer hybridization is far removed from the Dirac points. We test whether the top and bottom layers show the same cooling dynamics or whether the interlayer interaction and asymmetry affect the electronic response in this weakly hybridized part of the band structure. $T_e(t)$ was obtained by fitting the Fermi--Dirac cutoff in momentum-integrated TR-ARPES spectra. To test the robustness of the results, we used standard Gaussian-convoluted Fermi--Dirac distributions \cite{Johannsen:2013ab,Ulstrup:2014aa} and an extended model accounting for the finite angular resolution in the in-plane momentum direction perpendicular to the analyzer slit. Both approaches give consistent qualitative results (see Appendix), and we use the simpler model below. Fig.~\ref{fig:4} shows $T_e(t)$ for the top and bottom graphene layers; the corresponding result for QFMLG is shown in the Appendix. The two layers show very similar post-excitation cooling dynamics.

\begin{figure}[t]
\includegraphics[width=0.5\textwidth]{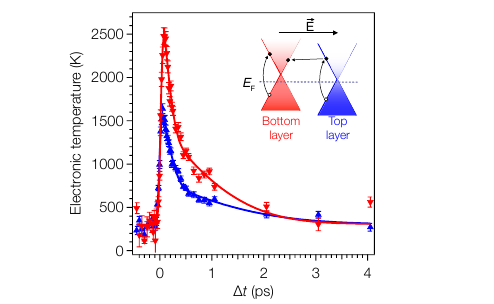}
\caption{(Color online) 
Time-dependent electronic temperature $T_e(t)$ for the top layer (blue) and bottom layer (red). Symbols are experimental data; solid lines are fits to the three-temperature model described in the text. The inset sketches the proposed ultrafast electron transfer from the top to the bottom layer in the built-in interlayer electric field $\vec{\mathrm{E}}$. The curved arrows indicate optical excitation, and the arrow between the cones indicates interlayer electron transfer.  
}
\label{fig:4}
\end{figure}

The most remarkable result from this comparison is that the bottom layer reaches a substantially higher peak temperature than the top layer, that is $\approx$ 2500~K compared with $\approx$ 1640~K. The large twist angle allows us to observe this asymmetry directly by measuring the electronic temperatures of the two layers independently. The observed temperature difference is unexpected for direct optical excitation of the two graphene layers. Comparable excitation of the two layers would be expected, or possibly a slightly reduced excitation of the bottom layer due to attenuation of the pump field in the top layer~\cite{Nair:2008aa}. The asymmetry therefore points to an ultrafast redistribution of the deposited electronic energy between the two layers. Their different doping levels break the layer symmetry and produce a built-in interlayer electric field of order $0.1$~V/nm (see Appendix). Under strongly non-equilibrium conditions immediately after photoexcitation, this field can bias interlayer transfer and thereby redistribute energy on an ultrafast time scale.
 
We investigate the electronic cooling using a three-temperature model similar to Ref.~\cite{Johannsen:2013ab}, in which the energy deposited into the electronic system is subsequently transferred to strongly coupled optical phonons and the remaining lattice. We fit the top and bottom $T_e(t)$ traces separately, allowing the electron--phonon coupling parameters $\lambda_1$ and $\lambda_2$, the fraction of strongly coupled optical phonons $R$, and the excitation strength to vary. The excitation is described by a Gaussian source term $S(t)/\beta$ with a FWHM of approximately 80~fs, where $\beta$ controls the effective deposited energy. We also account for the small difference in electronic heat capacity and $k_\mathrm{F}$ arising from the different doping levels. The resulting parameters are very similar for the two curves and for the QFMLG (see Appendix), with the main difference being the excitation strength $1/\beta$ (see Appendix). The fitted curves for the top and bottom layers are shown in Fig.~\ref{fig:4}. They reproduce the measured dynamics reasonably well.

Nearly identical cooling parameters but different excitation strengths suggest that the temperature asymmetry is established during the excitation and electron-thermalization stage. A substantial interlayer energy transfer persisting after excitation would generally modify the decay shapes, which is not observed. Ultrafast interlayer transfer on sub-100~fs time scales has been observed in van der Waals heterostructures, involving both charge transfer and interfacial energy transfer~\cite{Hong:2014aa,Dong:2023aa}. In our system, the different doping levels of the two graphene layers give rise to a built-in electrostatic potential difference, which could favor transfer in one direction under photoexcitation. This is illustrated by the sketch in the inset of Fig.~\ref{fig:4} . Such an ultrafast redistribution would be indistinguishable from a layer-dependent excitation strength in our model.
 
In conclusion, spatially resolved TR-ARPES reveals a spectroscopic signature consistent with the strongly bound ghost-Fano exciton, including its energy, lifetime, location in $k$ space, and spatial extent. The momentum separation of the layer-polarized Dirac cones further enables layer-resolved hot-carrier thermometry and reveals markedly different peak electronic temperatures in the two graphene sheets. We attribute this asymmetry to field-biased ultrafast interlayer energy redistribution, illustrating how twist and electrostatic asymmetry can expose and potentially control non-equilibrium interlayer dynamics in a homobilayer.

\section*{Acknowledgements}
We gratefully acknowledge valuable discussions with Matthijs Jansen and Stefan Mathias, as well as funding from Independent Research Fund Denmark  (Grant No. 4258-00002B) and Novo Nordisk Foundation (Grant number NNF23OC0085585).
The work was co-funded by the European
Union (ERC grant EXCITE with project number
101124619). Views and opinions expressed are however
those of the authors only and do not necessarily reflect
those of the European Union or the European Research
Council. Neither the European Union nor the granting
authority can be held responsible for them.

\section{Appendix}

\subsection{Experimental details}

\subsubsection{Sample fabrication}

\begin{figure}[b]
\includegraphics[width=0.55\textwidth]{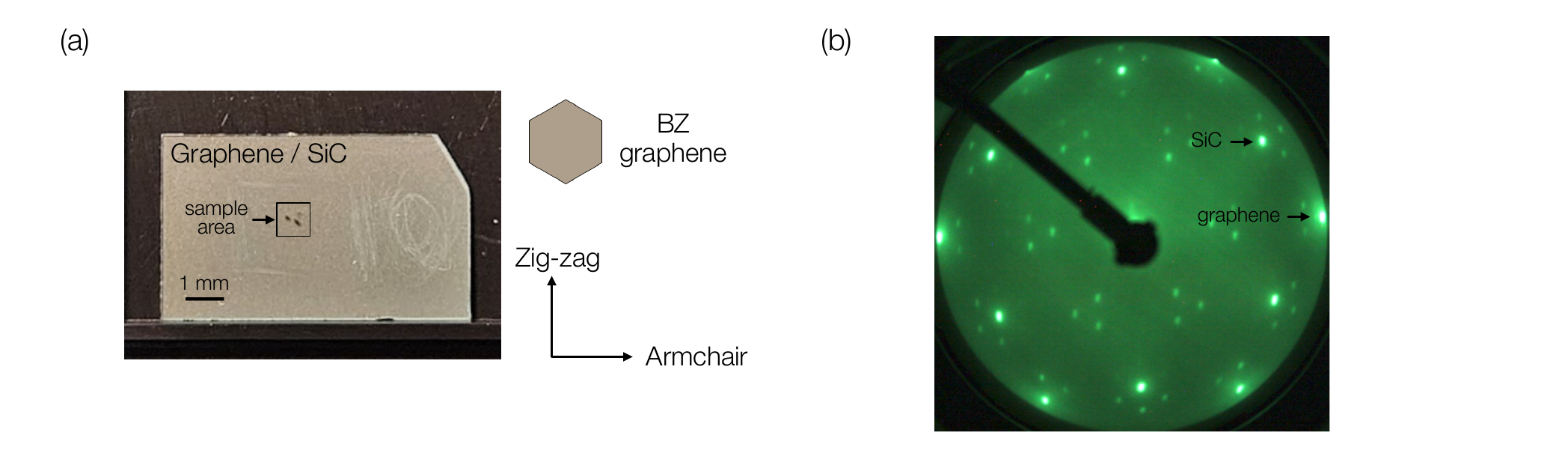}
\caption{(Color online) Characterization of the quasi-free-standing graphene/SiC substrate.
(a) Optical image of the substrate. On the right-hand side, the orientations of the graphene real-space lattice and Brillouin zone are sketched.
(b) Low-energy electron diffraction pattern measured at a kinetic energy of 91~eV on the substrate.}
\label{fig:S1}
\end{figure}

The tBLG samples were fabricated by combining surface preparation and dry-transfer techniques. The bottom graphene layer was prepared by thermal annealing and hydrogen intercalation of a SiC crystal, producing a mm-scale quasi-free-standing monolayer graphene sheet with weak interaction with the underlying SiC substrate [Fig.~\ref{fig:S1}(a)] \cite{Riedl:2009aa}. The top graphene layer was obtained by exfoliating a graphite single crystal onto a SiO$_2$/Si wafer. Graphene flakes with lateral dimensions of approximately $40\times40~\mu\mathrm{m}^2$ were identified by optical microscopy, picked up, and released onto the graphene/SiC substrate using a polycarbonate stamp. After transfer, polymer residues were removed by rinsing in chloroform, followed by annealing in a small flow of H$_2$/Ar gas at 220$^{\circ}$C for 2~h.

The armchair and zigzag directions of the bottom graphene were inferred from the low-energy electron diffraction (LEED) pattern of the bare graphene/SiC substrate. Fig.~\ref{fig:S1}(b) shows a representative LEED pattern. The two most intense sets of diffraction spots arise from the SiC and graphene lattices, while weaker replicas originate from the interface superlattice, consistent with Ref.~\cite{Razado-Colambo:2018aa}. This determines the orientation of the graphene real-space lattice shown in Fig.~\ref{fig:S1}(a). These directions were found to be essentially parallel to the edges of the as-diced SiC wafer and were used as a first reference for the twist-angle alignment.

The high-symmetry directions of the exfoliated graphene flakes were estimated from their straight edges in optical micrographs. These edges were assigned to either the armchair or zigzag direction and used as the second reference for the rotational alignment. Because the assignment of armchair and zigzag edges is ambiguous, this procedure determines the twist angle only modulo 30$^{\circ}$. Several monolayer flakes were therefore stacked on the same graphene/SiC substrate with a target twist angle of 10$^{\circ}$. A microscope image of the flake studied here is shown in Fig.~\ref{fig:1}(a). The little red square in the centre is an artifact from the microscope settings. The exact twist angle was subsequently determined from the positions of the Dirac cones in reciprocal space by ARPES, as described below.

\subsubsection{Twist angle estimated from ARPES}

\begin{figure}[t]
\includegraphics[width=0.45\textwidth]{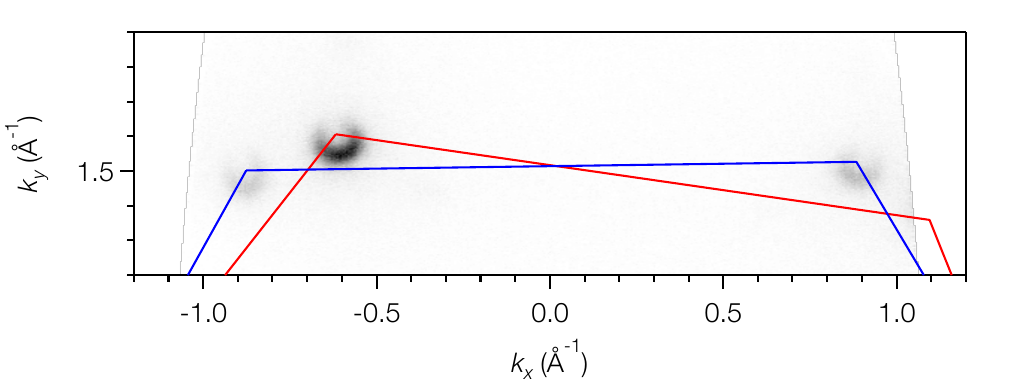}
\caption{(Color online) Characterization of the twist angle. Constant-energy map measured 200~meV below the Fermi energy on the tBLG sample at the SGM-4 beamline using 70~eV photons with linear horizontal polarization. Red (blue) lines mark part of the hexagonal Brillouin zone of the top (bottom) graphene layer.}
\label{fig:S2}
\end{figure}

The twist angle was determined after fabrication from high-resolution ARPES measurements performed at the SGM-4 beamline at ASTRID2, Aarhus \cite{Jones:2025aa}. The electronic structure was mapped in momentum space around the K points of both layers. A constant-energy map measured 200~meV below the Fermi energy is shown in Fig.~\ref{fig:S2}. The circular contours arise from cuts through the $\pi$ bands below the respective Dirac points. The most intense contour is associated with the top graphene layer, while the fainter contours are attributed to the bottom layer. We use the bottom-layer K--K' distance to calibrate the $k$-scaling, and then determine the orientations of the top-layer Brillouin zone (BZ) (red) and bottom-layer BZ (blue). Their relative rotation gives a twist angle of $9^{\circ}$, with an estimated uncertainty of approximately $1^{\circ}$.

\subsubsection{Fluence calculation during time-resolved ARPES experiments}

The laser repetition rate was 100~kHz and the pump power measured before the vacuum viewport was 9.7~mW, corresponding to an energy of 97~nJ per pulse. Accounting for a reflection coefficient of 0.95 for the last focusing mirror gives a pulse energy of approximately 92~nJ at the sample. The dimensions of the pump spot were extracted from the camera image of the spot on a yttrium aluminum garnet scintillator crystal at the sample position. We approximated its profile as Gaussian with estimated $1/e^2$ diameters of $d_x \approx 0.17$~mm and $d_y \approx 0.18$~mm. For an elliptical Gaussian beam, the peak fluence at normal incidence is
\begin{equation}
F_0=\frac{8E_{\mathrm{pulse}}}{\pi d_x d_y}.
\end{equation}
At an incidence angle $\theta=10^{\circ}$ from the surface normal, the increased footprint on the sample reduces the peak fluence by a factor of $\cos\theta$. This gives
\begin{equation}
F_{\mathrm{peak}}=F_0\cos\theta\approx0.75~\mathrm{mJ/cm^2}.
\end{equation}

\subsubsection{Estimate of the interlayer electric field}

The Dirac points of the top and bottom graphene layers are located 63 and 93~meV above the Fermi energy, respectively, corresponding to an energy offset of approximately 30~meV. Interpreting this offset predominantly as an electrostatic potential difference between the two graphene sheets gives $\Delta V\approx30$~mV. Using an interlayer separation of $d\approx3.4$~\AA, the corresponding field is
\begin{equation}
    E_{\perp}\approx\frac{\Delta V}{d}\approx0.09~\mathrm{V/nm}.
\end{equation}
This is an order-of-magnitude estimate.

\section{Tight-binding model}

We describe the electronic structure near the intersection of the two Dirac cones using a minimal four-band tight-binding model. In the basis of the two sublattices in each graphene sheet, the bilayer Hamiltonian is written as
\begin{equation}
    h_{\mathrm{tBLG}}(\mathbf{k})=
    \begin{pmatrix}
        h_{\mathrm{ML}}(\mathbf{k}) & \hat{T}\\
        \hat{T}^{\dagger} & h_{\mathrm{ML}}(\mathbf{k}_{\theta})
    \end{pmatrix},
\end{equation}
where $\mathbf{k}_{\theta}=\hat{M}_{\theta}\mathbf{k}$ and $\hat{M}_{\theta}$ is the two-dimensional rotation matrix for the twist angle $\theta$.

For each graphene layer we use the nearest-neighbor monolayer Hamiltonian
\begin{equation}
    h_{\mathrm{ML}}(\mathbf{k})=-t
    \begin{pmatrix}
        0 & f(\mathbf{k})\\
        f^{*}(\mathbf{k}) & 0
    \end{pmatrix},
\end{equation}
with
\begin{equation}
    f(\mathbf{k})=\exp\!\left(-i\mathbf{k}\cdot\mathbf{d}_1\right)+\exp\!\left(-i\mathbf{k}\cdot\mathbf{d}_2\right)
                   +\exp\!\left(-i\mathbf{k}\cdot\mathbf{d}_3\right),
\end{equation}
where $\mathbf{d}_1$, $\mathbf{d}_2$ and $\mathbf{d}_3$ are the real-space vectors connecting a carbon atom to its nearest neighbors. The parameter $t$ is the intralayer nearest-neighbor hopping integral.

The two layers are coupled phenomenologically through
\begin{equation}
    \hat{T}=
    \begin{pmatrix}
        0 & \Delta\\
        \Delta & 0
    \end{pmatrix}.
\end{equation}
The model parameters are the twist angle $\theta$, the intralayer hopping $t$, and the interlayer coupling $\Delta$. We use $\theta=9^{\circ}$ from the ARPES determination above and $t=3$~(2.83)~eV for top (bottom) graphene, consistent with the measured band dispersion in each layer. The parameter $\Delta$ controls the avoided crossing of the two cones; a fit to the energy-distribution curve at the midpoint between the two K points gives a hybridization gap of approximately 360~meV and therefore $\Delta=180$~meV.
The model assumes particle--hole symmetry, which is not necessarily fulfilled over a large energy range in graphene. For the twist-induced vHs in the occupied and unoccupied states considered here, this symmetry breaking is sufficiently small to be neglected~\cite{Brihuega:2012aa}.

\section{Layer-dependent electronic temperature}

We determine the electronic-temperature dynamics of the two graphene layers following the approach of Refs.~\cite{Johannsen:2013ab,Ulstrup:2014aa}. First, the electronic temperature $T_e(t)$ of each layer is extracted from the Fermi edge of momentum-integrated TR-ARPES spectra. The resulting temperature traces are then analyzed independently with the three-temperature model described below.

\subsubsection{Temperature fits}

\begin{figure}[h!]
\includegraphics[width=0.35\textwidth]{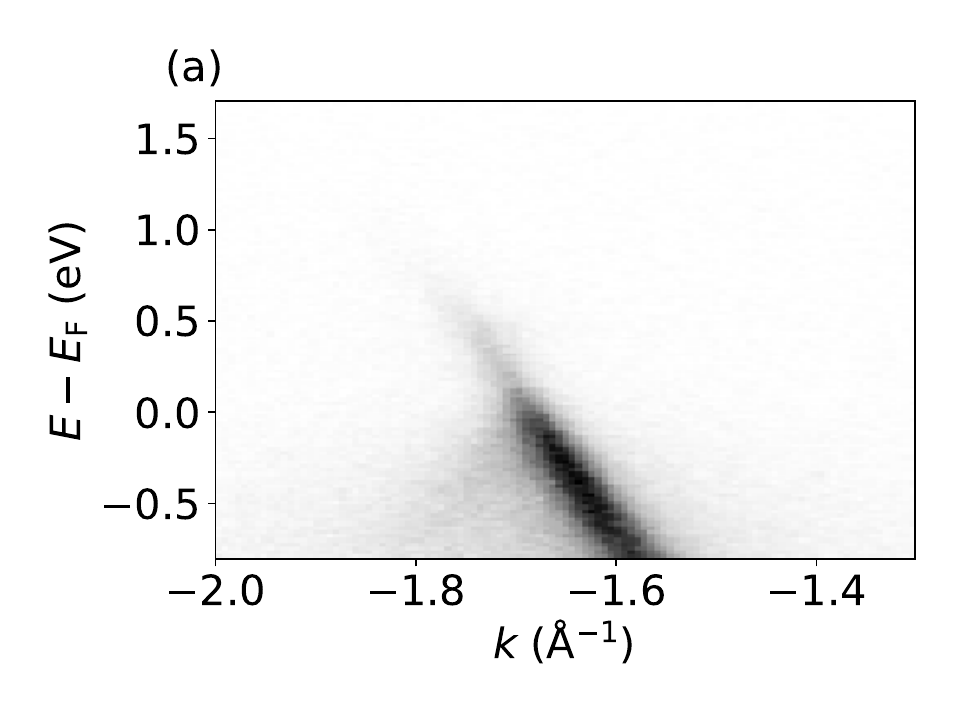}
\includegraphics[width=0.35\textwidth]{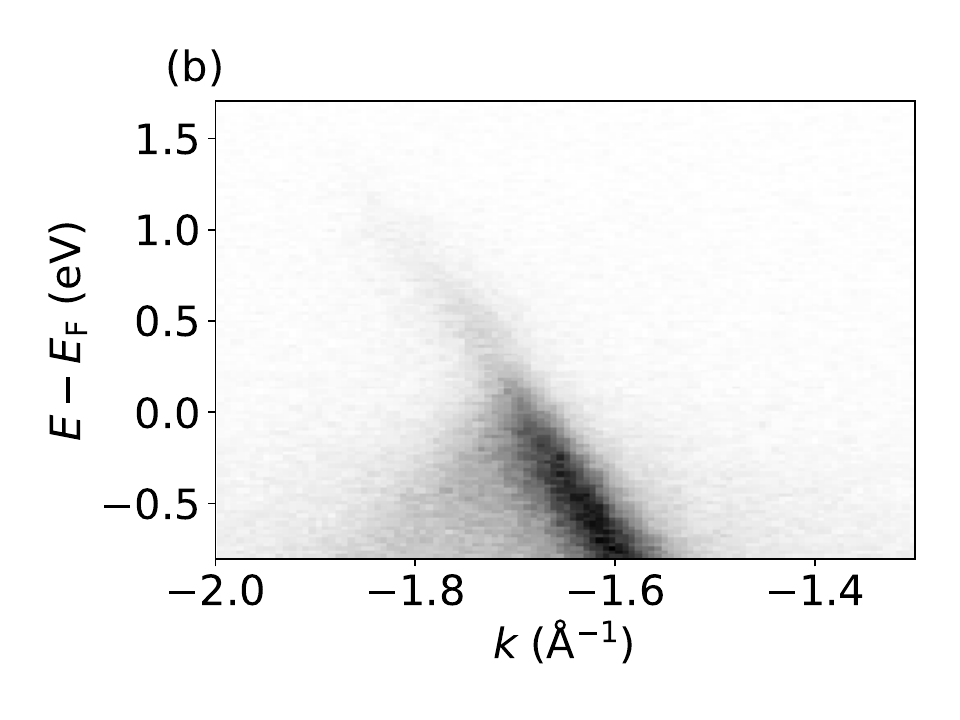}
\caption{ARPES measurements of the tBLG sample near K$_\mathrm{t}$ (a) and K$_\mathrm{b}$ (b) at peak excitation ($\Delta t = 50$~fs). The displayed ranges correspond to the regions used in the electronic-temperature analysis.}
\label{fig:S3}
\end{figure}

\begin{figure}[h!]
\includegraphics[width=0.35\textwidth]{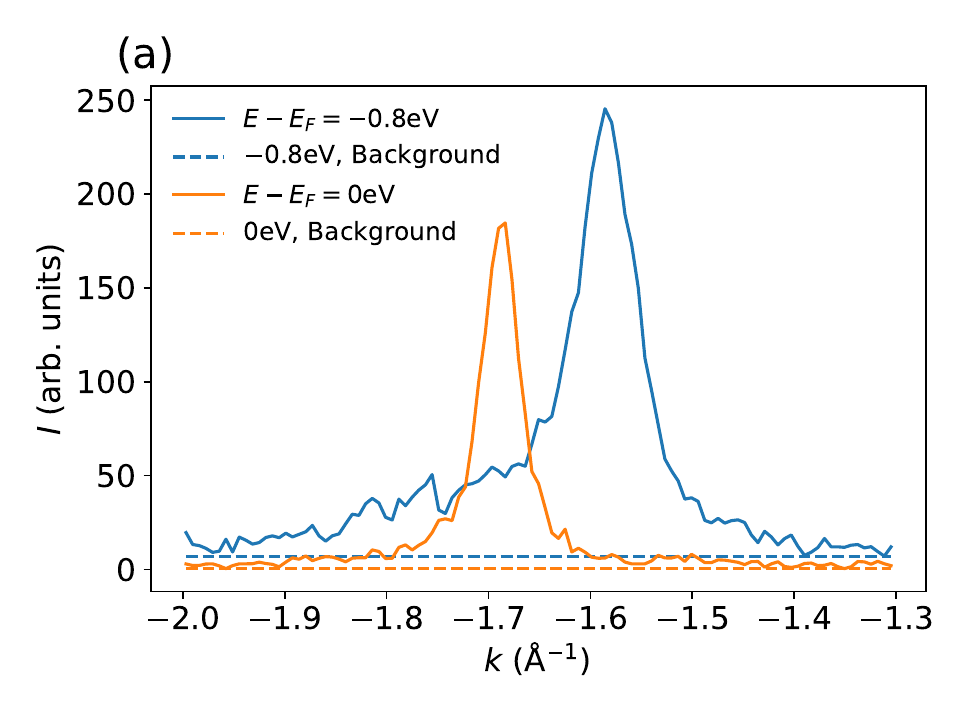}
\includegraphics[width=0.35\textwidth]{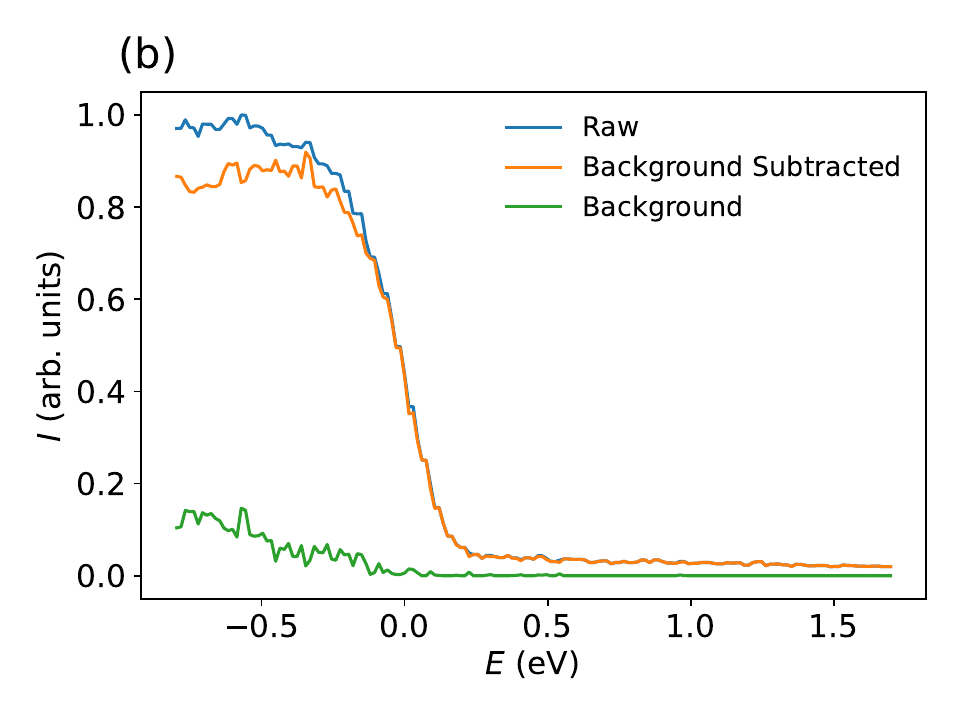}
\caption{Background subtraction used for the temperature analysis. (a) Representative MDCs (solid lines) at two energies together with the estimated background contributions (dashed lines). (b) Momentum-integrated EDCs at the same delay before and after background subtraction; the estimated background is also shown. The data was measured near K$_\mathrm{t}$ at $T_e=300$ K.}
\label{fig:S4}
\end{figure}

For each layer we selected a region of interest (ROI) around K$_\mathrm{t}$ or K$_\mathrm{b}$ with the same energy and momentum range. The two ROIs are illustrated in Fig.~\ref{fig:S3}. While the ROI is relatively wide in momentum, a weak residual photoemission intensity remains at the edges of the selected $k$ range. We estimate this contribution at each energy from the minimum intensity in the corresponding momentum distribution curve (MDC) and subtract it before integrating over momentum, as illustrated in Fig.~\ref{fig:S4}. Since this procedure is sensitive to noise and tends to underestimate the background, a residual linear background is retained in the subsequent fits.

This procedure differs slightly from the analysis in Ref.~\cite{Ulstrup:2014aa}, where the MDCs were fitted before extracting the energy distribution. We verified that fitting the MDCs instead of directly integrating the background-corrected intensity has little influence on the resulting energy distribution curve (EDC), consistent with Ref.~\cite{Ulstrup:2014aa}.

\begin{figure}[h!]
\includegraphics[width=0.35\textwidth]{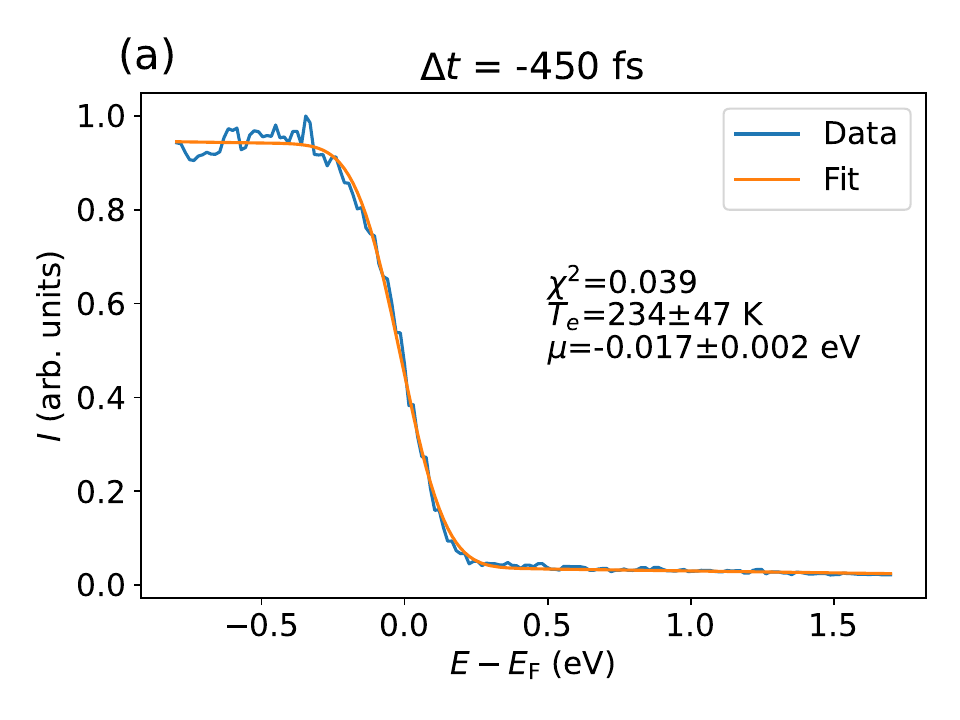}
\includegraphics[width=0.35\textwidth]{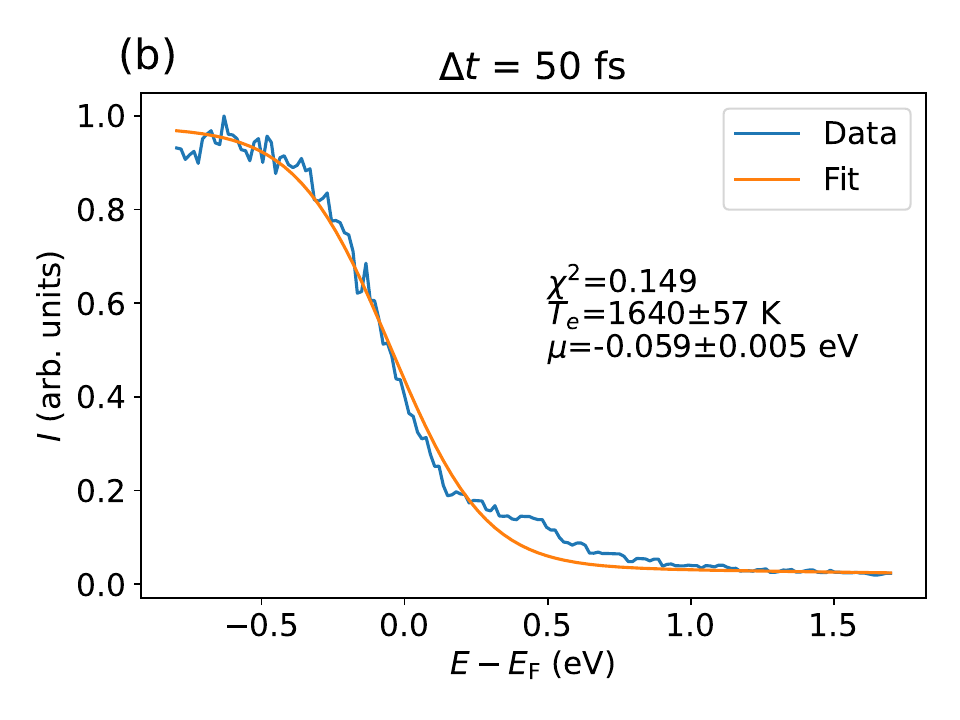}
\caption{Representative EDC fits using the simple Fermi-edge model at a negative time delay (a) and at peak excitation (b).}
\label{fig:S5}
\end{figure}

The time-dependent, momentum-integrated EDCs are normalized to a maximum intensity of one and fitted with
\begin{equation}
    F(E)=A\left[f(E;T_e,\mu)\otimes G_{\sigma_E}(E)\right]
    +a(E-E_{\mathrm F})+b,
\end{equation}
where $f(E;T_e,\mu)$ is the Fermi--Dirac distribution and $G_{\sigma_E}(E)\propto\exp[-E^2/(2\sigma_E^2)]$ represents the experimental energy broadening. The remaining parameters are an overall amplitude $A$ and the slope $a$ and intercept $b$ of a linear background.

The fits are performed in two stages. First, the average EDC acquired before $t_0$ is fitted separately for each layer with $T_e$ fixed to 300~K and all other parameters free. The resulting $\sigma_E$, $a$, and $b$ are then fixed. In the second run, only $A$, $\mu$, and $T_e$ are allowed to vary at each  time delay. Allowing $\mu$ to vary accounts phenomenologically for small energy shifts in the spectra \cite{Zhou:2005ab}. Representative fits are shown in Fig.~\ref{fig:S5}. The model describes the low-temperature spectra well. Near peak excitation, however, additional intensity is observed as a ``bump'' above the Fermi level that is not fully captured by the simple Fermi-edge model.

\subsubsection{Finite slit model}

One possible instrumental contribution to the ``bump''-shaped excess intensity above $E_{\mathrm F}$ [Fig.~\ref{fig:S5}(b)] is the finite angular acceptance perpendicular to the long axis of the analyzer slit. We therefore tested a model that integrates a linearly dispersing Dirac cone over both in-plane momentum directions.

Assuming a Lorentzian intrinsic spectral broadening with half width at half maximum $\Gamma$, a Gaussian acceptance function in the direction perpendicular to the slit, and a linear Dirac dispersion, the momentum-integrated spectral weight before multiplication by the Fermi--Dirac distribution can be written as
\begin{equation}
\begin{aligned}
S(E) \propto {} & \sum_{s=\pm1}
\int_{-k_{\parallel,0}}^{k_{\parallel,0}}
\int_{-k_{\perp,0}}^{k_{\perp,0}}
\\
& \times
\frac{\exp[-k_{\perp}^{2}/(2\sigma_{\perp}^{2})]}
{\Gamma^{2}+\left[E-E_D-s\hbar v_{\mathrm F}
\sqrt{k_{\parallel}^{2}+k_{\perp}^{2}}\right]^{2}}
\,dk_{\perp}\,dk_{\parallel}.
\end{aligned}
\label{FiniteSlitModel}
\end{equation}

where $s=-1$ and $+1$ denote the lower and upper Dirac branches, respectively, $E_D$ is the Dirac-point energy, and $\sigma_{\perp}$ parametrizes the perpendicular momentum acceptance. The limits $k_{\parallel,0}$ and $k_{\perp,0}$ describe the momentum range accepted by the analyzer.

Since the perpendicular momentum acceptance is predominantly determined by the Gaussian factor and the relevant momentum range is small compared with $k_{\parallel,0}$, we approximate the rectangular integration domain by a circular one with radial cutoff $k_{\parallel,0}$. Introducing polar coordinates and performing the angular integration analytically gives
\begin{equation}
S(E)\propto
\sum_{s=\pm1}
\int_0^{k_{\parallel,0}}
\frac{k\,\exp[-k^2/(4\sigma_\perp^2)]
I_0[k^2/(4\sigma_\perp^2)]}
{\Gamma^2+(E-E_D-s\hbar v_{\mathrm F}k)^2}\,dk ,
\end{equation}
where $I_0$ is the zeroth-order modified Bessel function of the first kind \cite{NIST:DLMF}. The remaining radial integral is evaluated numerically. The full fitting function is obtained by multiplying $S(E)$ by the Fermi--Dirac distribution, convolving with the experimental energy-resolution function, and adding a linear background,

\begin{equation}
\begin{aligned}
I(E)=A\Bigg[
& f(E;T_e,\mu)
\sum_{s=\pm1}
\\
& \times
\int_{0}^{k_{\parallel,0}}
\frac{k\,\exp[-k^2/(4\sigma_{\perp}^2)]
I_0[k^2/(4\sigma_{\perp}^2)]}
{\Gamma^2+(E-E_D-s\hbar v_{\mathrm F}k)^2}\,dk
\Bigg]
\\
& \otimes G_{\sigma_E}(E)
+a(E-E_{\mathrm F})+b .
\end{aligned}
\end{equation}

The Gaussian energy broadening and the linear background were fixed from the pre-pump EDCs, as for the simple model. The parameters $\Gamma$ and $\sigma_{\perp}$ were estimated from the MDC width and the analyzer-slit acceptance rather than fitted freely. The remaining parameters $A$, $\mu$, and $T_e$ were then obtained independently at each delay.

\begin{figure}[h!]
\includegraphics[width=0.35\textwidth]{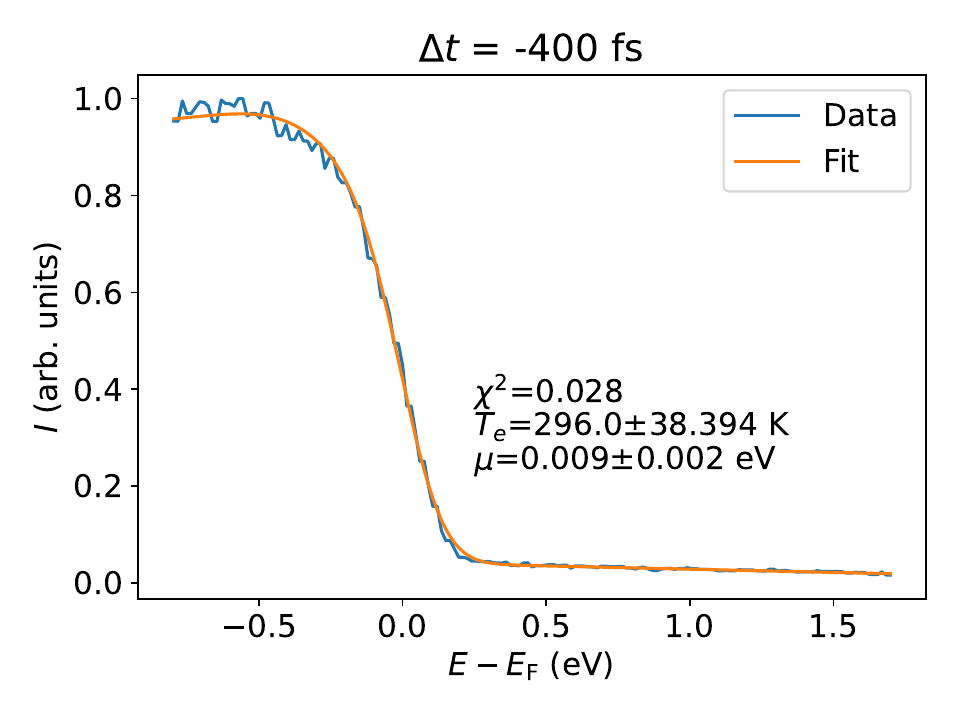}
\includegraphics[width=0.35\textwidth]{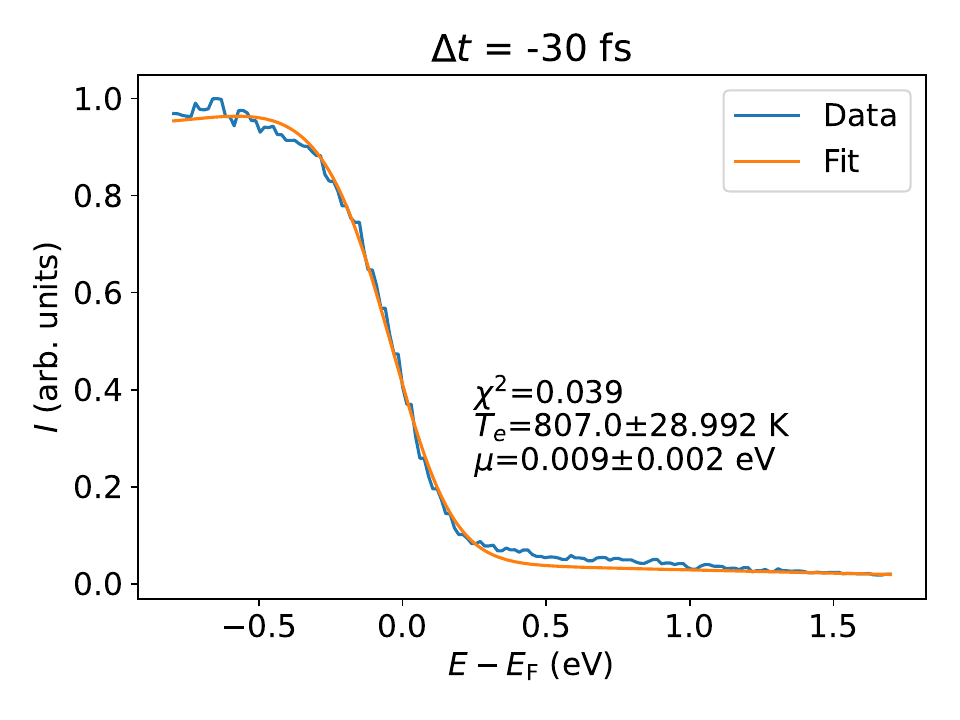}
\includegraphics[width=0.35\textwidth]{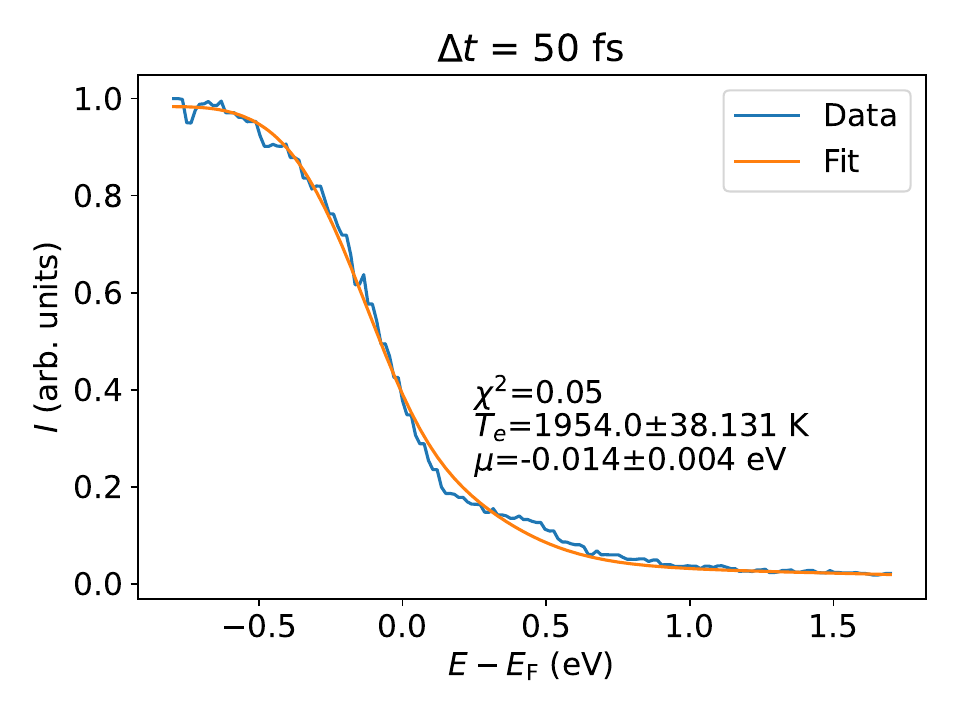}
\includegraphics[width=0.35\textwidth]{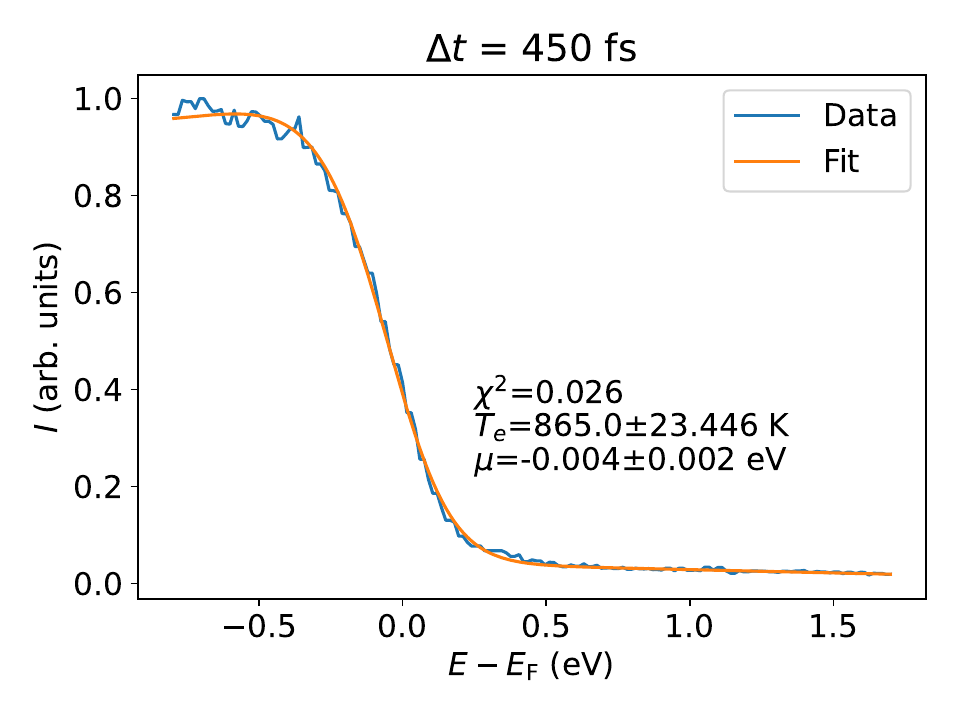}
\includegraphics[width=0.35\textwidth]{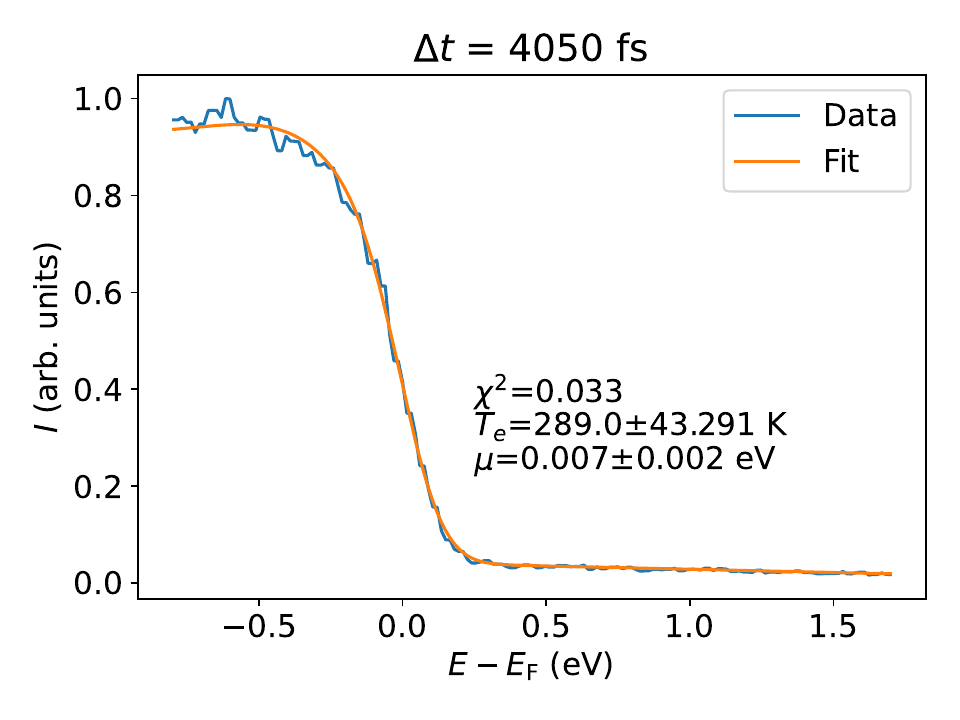}
\caption{Representative EDC fits using the finite-slit Fermi-edge model at five pump--probe delays before, around, and after peak excitation.}
\label{fig:S6}
\end{figure}

\begin{figure}[h!]
\includegraphics[width=0.35\textwidth]{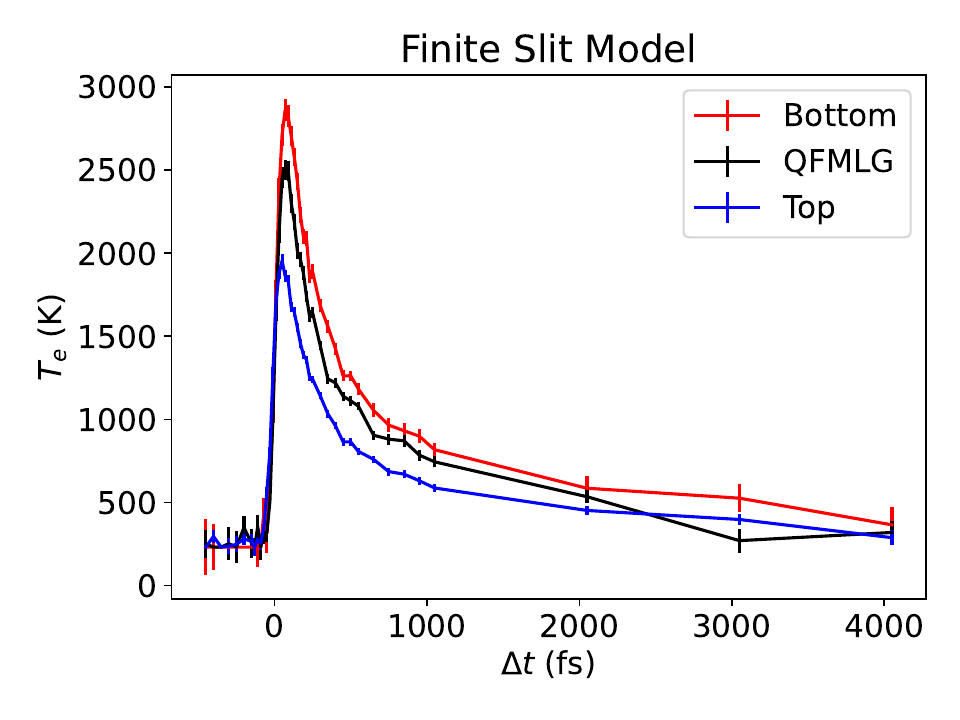}
\caption{Electronic-temperature traces obtained with the finite-slit model. }
\label{fig:S7}
\end{figure}

Representative fits are shown in Fig.~\ref{fig:S6}. The finite-slit model describes the spectra well long before and after excitation but still shows systematic deviations while the electronic system is being heated at $\Delta t = 50$~fs. Spectra with comparable fitted temperatures on the rising and falling sides of the transient have different line shapes. The excess intensity is therefore not explained solely by the finite angular acceptance: for fixed $\Gamma$ and $\sigma_{\perp}$, an instrumental slit effect would depend on temperature but not on whether the system is heating or cooling. As shown in Fig.~\ref{fig:S7}, the finite-slit model gives somewhat higher temperatures than the simple analysis throughout, but the qualitative layer-dependent dynamics is unchanged. We use the simple Fermi-edge model for the analysis in the main paper.

\subsubsection{Three-temperature model}

We use the phenomenological three-temperature model introduced in Ref.~\cite{Johannsen:2013ab}. For each graphene sheet, the model contains three coupled thermal reservoirs: electrons at temperature $T_e$, strongly coupled optical phonons at temperature $T_p$, and the remaining lattice degrees of freedom, dominated by acoustic phonons, at temperature $T_l$. There are no coupling terms between the top- and bottom-layer models.

For each layer, the temperature dynamics are described by
\begin{align}
\frac{\partial T_e}{\partial t} &= \frac{S(t)}{\beta}
-J_{\mathrm{op}}-J_{\mathrm{sc}}-J_{\mathrm{ac}}, \\
\frac{\partial T_p}{\partial t} &= \frac{C_e}{C_p}J_{\mathrm{op}}, \\
\frac{\partial T_l}{\partial t} &= \frac{C_e}{C_l}\left(J_{\mathrm{sc}}+J_{\mathrm{ac}}\right),
\label{DE}
\end{align}
with
\begin{align}
J_{\mathrm{op}} &=
\frac{\pi\lambda_1 g[\mu(T_e)]\Omega_{\mathrm{Ein}}^3}{\hbar}
\frac{n_e-n_p}{C_e},\\
J_{\mathrm{sc}} &=
9.62\frac{\lambda_2 g[\mu(T_e)]k_{\mathrm B}^3}{\hbar k_{\mathrm F}l}
\frac{T_e^3-T_l^3}{C_e},\\
J_{\mathrm{ac}} &=
\pi\lambda_2\hbar g[\mu(T_e)]k_{\mathrm F}^2v_s^2k_{\mathrm B}
\frac{T_e-T_l}{C_e}.
\end{align}
Here $C_e$, $C_p$, and $C_l$ are the electronic, optical-phonon, and remaining lattice heat capacities; $\lambda_1$ and $\lambda_2$ are the electron--phonon coupling constants for optical and acoustic phonons, respectively; $v_s$ is the sound velocity; $k_{\mathrm F}$ is the Fermi wave vector measured relative to K; and $l$ is the defect-scattering mean free path. The optical phonons are represented by an Einstein mode with energy $\Omega_{\mathrm{Ein}}=0.2$~eV. The quantities
\begin{equation}
 n_{e,p}=\left[\exp\!\left(\frac{\Omega_{\mathrm{Ein}}}{k_{\mathrm B}T_{e,p}}\right)-1\right]^{-1}
\end{equation}
are the Bose occupation factors of this mode evaluated at the electronic and optical-phonon temperatures, respectively. The supercollision and momentum-conserving acoustic-phonon terms follow Refs.~\cite{Graham:2013aa,Betz:2013aa,Johannsen:2013ab}.

The pump term $S(t)$ is represented by a Gaussian centered at $t_0$. Writing its full width at half maximum as $w$ and the incident pump fluence as $F_P$, we use
\begin{equation}
S(t)=F_P\,\frac{2\sqrt{\ln2}}{w\sqrt{\pi}}
\exp\!\left[-4\ln2\frac{(t-t_0)^2}{w^2}\right].
\end{equation}
The scaling parameter $\beta$ has the dimensions of an areal heat capacity and controls the magnitude of the pump-induced temperature rise. It should not be interpreted as the equilibrium electronic heat capacity or as an absorbed fraction of the pump fluence. As in Ref.~\cite{Johannsen:2013ab}, it is used phenomenologically because the electronic system is not in thermal equilibrium during the earliest part of the excitation. Only the ratio $F_P/\beta$ enters the source term, so various experimental uncertainties can be incorporated into the effective value of $\beta$.

For the supercollision term, $k_{\mathrm F}$ and $l$ are fixed from the high-resolution ARPES data. The present fits used $k_{\mathrm F}=0.02$~\AA$^{-1}$ and $l\approx1/(0.06~$\AA$^{-1})=16.7$~\AA~ for the bottom layer and $k_{\mathrm F}=0.013$~\AA$^{-1}$ and $l\approx1/(0.05~$\AA$^{-1})=20$~\AA ~for the top layer.

\begin{figure}[h!]
\includegraphics[width=0.35\textwidth]{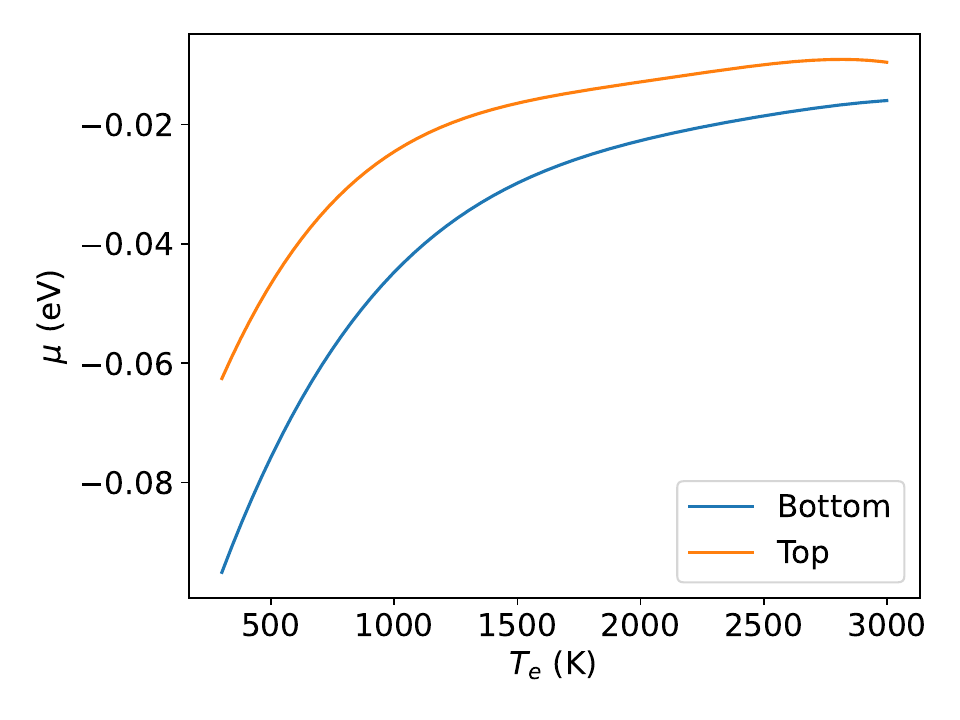}
\includegraphics[width=0.35\textwidth]{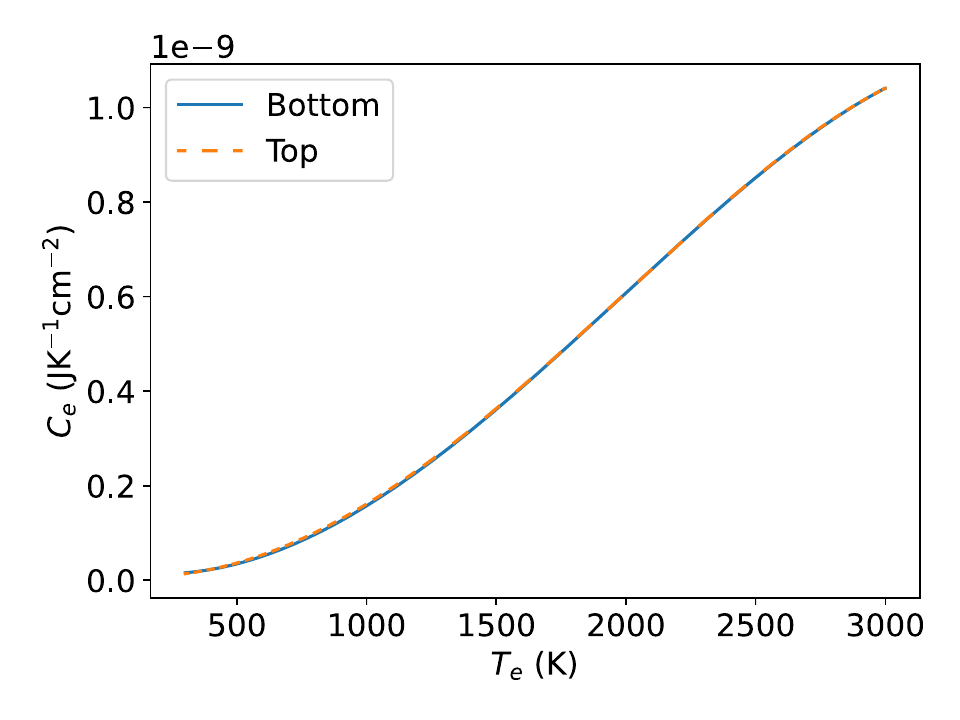}
\caption{Temperature-dependent chemical potential $\mu(T_e)$ (left) and electronic heat capacity $C_e(T_e)$ (right) used in the three-temperature model for the top and bottom graphene layers.}
\label{fig:S8}
\end{figure}

The electronic heat capacity is calculated separately for the two layers
to account for their different equilibrium doping. Taking the Dirac point
as the zero of energy, we use the graphene density of states including
spin and valley degeneracy,
\begin{equation}
    g(E)=\frac{2|E|}{\pi(\hbar v_F)^2},
\end{equation}
with $v_F=10^6$~m/s.

The linear Dirac dispersion is used over a finite energy range
$[-E_c,E_c]$ around the Dirac point. The cutoff $E_c$ is chosen
sufficiently large that states near the integration boundaries do not
change their occupation over the temperature range considered and the
resulting $\mu(T)$ and $C_e(T)$ are insensitive to its precise value. At each
temperature, the chemical potential is determined by conserving the
electron density within this range,
\begin{equation}
    n_0=\int_{-E_c}^{E_c}
    g(E)f(E;\mu,T)\,dE .
\label{CarrierConc}
\end{equation}
The corresponding electronic energy density is
\begin{equation}
    U_e(T)=\int_{-E_c}^{E_c}
    g(E)E f(E;\mu,T)\,dE ,
\end{equation}
and the electronic heat capacity at fixed carrier density is obtained
numerically as
$C_e=(\partial U_e/\partial T)_{n_0}$.
For each layer, $n_0$ is fixed by evaluating Eq.~(\ref{CarrierConc})
at the measured equilibrium chemical potential and the equilibrium
temperature.
The resulting $\mu(T_e)$ and $C_e(T_e)$ are shown in Fig.~\ref{fig:S8} and are interpolated during the numerical solution of the rate equations.

The heat capacity of the strongly coupled optical phonons is represented by an Einstein mode. Only a fraction $R$ of the available optical vibrations is assumed to participate efficiently in the energy exchange with the electrons. Expressed per unit area,
\begin{equation}
    C_p=R n_{\mathrm{uc}}\Omega_{\mathrm{Ein}}\frac{\partial n_p}{\partial T_p},
\end{equation}
where $n_{\mathrm{uc}}$ is the areal density of graphene unit cells. The remaining lattice heat capacity is estimated from the total heat capacity of graphene in Ref.~\cite{Benedict:1996aa} as $C_l=C_{\mathrm{tot}}-C_p$.

\subsubsection{Electronic-temperature fit results}

The three-temperature equations are solved separately for the top and bottom layers. For each layer, the fitted parameters are $\lambda_1$, $\lambda_2$, $R$, $\beta$, the pump width $w$, and the temporal offset $t_0$. The fitted values for the top (bottom) layer are $\lambda_1=0.20$ (0.33), $\lambda_2=0.012$ (0.015), $R=0.025$ (0.030), $\beta=4.4\times10^{-7}$ ($2.2\times10^{-7}$)~J~m$^{-2}$~K$^{-1}$, $w=76$ (86)~fs, and $t_0=-14$ ($9$)~fs.

The parameters are strongly correlated: appreciable changes in an individual coupling parameter can be compensated by changes in the others without substantially worsening the fit. We therefore do not regard the modest differences in $\lambda_1$, $\lambda_2$, $R$, $w$, or $t_0$ between the two independent fits as significant. The substantially different source-scaling parameter $\beta$, however, is required to reproduce the different peak electronic temperatures. This is consistent with the interpretation in the main text that the two layers cool in essentially the same way after excitation but receive different effective electronic energy during excitation and thermalization. No explicit interlayer coupling term is required to describe the subsequent cooling.

\begin{figure}[h!]
\includegraphics[width=0.35\textwidth]{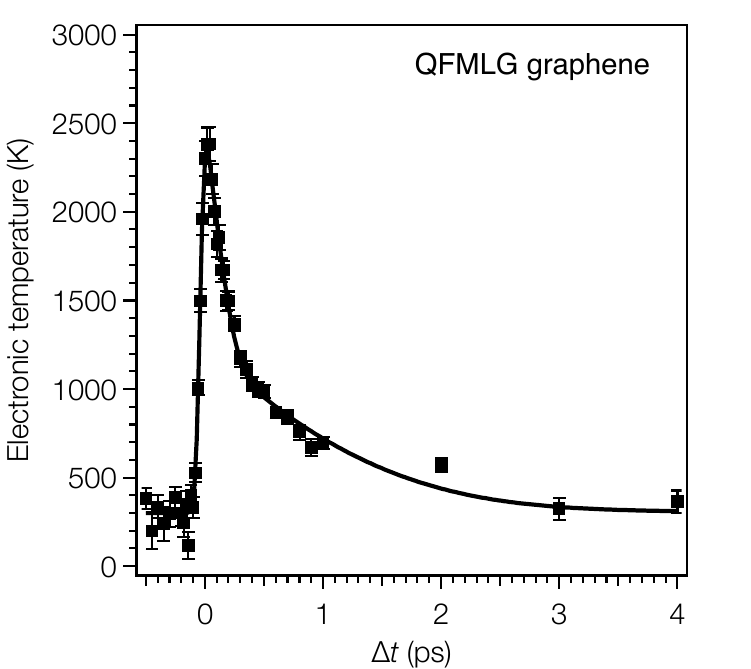}
\caption{Electronic temperature as a function of pump--probe delay for the uncovered quasi-free-standing monolayer graphene region (markers), together with the three-temperature-model fit (solid line).}
\label{fig:S9}
\end{figure}

Fig.~\ref{fig:S9} shows the corresponding three-temperature-model fit to the electronic temperature measured on the uncovered quasi-free-standing monolayer graphene region.

%

\end{document}